\documentclass[aps,twocolumn,prl,floats,amssymb]{revtex4}

\RequirePackage{xspace} 

\usepackage{epsfig}
\usepackage{graphicx}
\usepackage{color}

\def\meas{\textrm{\tiny meas}}
\def\beq{\begin{equation}}        
\def\eeq{\end{equation}}
\def\bea{\begin{eqnarray}}
\def\eea{\end{eqnarray}}

\def\nn{\nonumber}
\def\sss{\scriptscriptstyle}

\def\bd{B_d^0}
\def\bdbar{{\overline{B_d^0}}}

\def\barp{{\raise.35ex\hbox{${\sss (}$}}---{\raise.35ex\hbox{${\sss )}$}}}
\def\bdbarp{\hbox{$B_d$\kern-1.4em\raise1.4ex\hbox{\barp}}}
\def\bsbarp{\hbox{$B_s$\kern-1.4em\raise1.4ex\hbox{\barp}}}

\def\roughly#1{\mathrel{\raise.3ex\hbox{$#1$\kern-.75em\lower1ex\hbox{$\sim$}}}}
\def\barpk{{\raise.35ex\hbox{${\sss (}$}}--{\raise.35ex\hbox{${\sss )}$}}}
\def\bbarp{\hbox{$B$\kern-0.9em\raise1.4ex\hbox{\barpk}}}

\def\adir00{{a_{\sss dir}^{00}}}

\def\B00{B^{00}}
\def\Bp0{B^{+0}}

\begin{document}

\title{ Has New Physics already been seen in $\mathbf{B_d}$ meson decays? }
\author{Rahul Sinha}
\author{Basudha Misra}
\affiliation{The Institute of Mathematical Sciences, Taramani, Chennai
  600113, India} 
\author{Wei-Shu Hou}
\affiliation{Department of Physics, National Taiwan University,
  Taipei, Taiwan 106, R.O.C. } 

%
%

%
%
\begin{abstract}
  We show using a model independent approach that the deviation in
  measured $\bd-\bdbar$ mixing phase caused by pollution from another
  amplitude, within the Standard Model, is always less in magnitude, and
  has the same sign as the weak phase of the polluting amplitude.  The
  exception is to have large destructive interference between the two
  amplitudes.  We demonstrate that any deviation larger than a few
  degrees is only possible if the observed decay rate results from
  fine tuned cancellations between significantly larger quark level
  amplitudes.  These simple observations have very significant
  consequences for signals of New Physics.
\end{abstract}

\maketitle

The $\bd-\bdbar$ mixing phase $\beta/\phi_1$ measured in $b\to
c\bar{c}s$ and $b\to s\bar{q}q$ (where $q=u,d,s$) modes is found to
differ at $2.6\sigma$\cite{HFAG}. It has been speculated that this
discrepancy is a signal of New Physics (NP). Within the Standard Model
(SM) the amplitude for modes involving $b\to s$ transitions, get
contributions from two amplitudes with different weak phases.  Unless
the contribution from one of the amplitudes is negligible, one would
expect some discrepancy between the various measurements.
Estimates~\cite{other-work} of this discrepancy using hadronic
assumptions have indicated that the sign of the discrepancy within SM
is opposite to the observed value. Nevertheless, one wonders whether
large rescattering effects can then somehow cause the observed
discrepancy that is speculated to be NP.  Indeed, convincing arguments
regarding the nature of this discrepancy are lacking. In light of this
uncertainty, the relevant question is ``under what conditions can this
discrepancy be regarded as an unambiguous signal of NP?''  In this
letter we seek to answer this question using a completely model
independent approach.

We show that the deviation in the measured $\bd-\bdbar$ mixing phase
caused by pollution from another amplitude, within the SM, is not only
always less than the weak phase of the polluting amplitude, but also
always has the same sign as the weak phase of the polluting amplitude.
The only exception is to have large destructive interference between
the two amplitudes.  Without making any hadronic model based
assumptions, we examine the conditions under which such a destructive
interference is possible within the SM.  We find that any deviation
larger than a few degrees is only possible, if the observed branching
ratios result from fine tuned cancellations between amplitudes whose
squares are at least an order of magnitude larger than the branching ratios
themselves.

The most general amplitude for $b\to s$ transition modes may be
written as
\begin{equation}
  \label{eq:amp-0}
  A^{b\to s}\!\!={\cal A}_u e^{i\delta_u}v_u + {\cal A}_c e^{i\delta_c}v_c +
  {\cal A}_t e^{i\delta_t}v_t,
\end{equation}
where ${\cal A}_j$ and $\delta_j$ are the amplitude and strong phases
respectively for the three quark level contributions to the
transition; $v_j=V_{jb}^* V_{js}$, where $j$ is either $u$, $c$ or $t$ and
$V_{jk}$ are the elements of the CKM matrix.  The unitarity of CKM,
namely $v_u+v_c+v_t=0$ allows us to express the amplitude in terms of
only two contributions, which are distinguished by their weak phases.
In SM $v_c$ is real at least to order $\mathcal{O}(\lambda^6)$ in the
Wolfenstein parameterization \cite{Wolfenstein:1983yz}, whereas
$v_u=A\lambda^4(\rho+i\eta)$ and
$v_t=-A\lambda^2+A(\frac{1}{2}-\rho-i\eta) \lambda^4+
\mathcal{O}(\lambda^6)$.  The weak phase of $v_t$ referred to as
$\beta_s$ is very small
($\beta_s={1.045^o}^{+0.061^o}_{-0.057^o}$~\cite{CKMfit} ) as it
appears at $\mathcal{O}(\lambda^4)$. $v_u$ on the other hand involves
a much larger phase $\gamma$, but is suppressed by an additional factor
$\lambda^2$.

The amplitude can be parameterized in terms of
only one weak phase either by eliminating $v_t$ or $v_u$. It is
customary to eliminate $v_t$ and express the amplitude in terms of
$\gamma$ as:
\begin{equation}
  \label{eq:amp-2}
  A^{b\to s}\!\!= ({\cal A}_c e^{i\delta_c}-{\cal A}_t e^{i\delta_t})v_c +
  ({\cal A}_u e^{i\delta_u}-{\cal A}_t e^{i\delta_t})v_u~.
\end{equation}
This can be re-expressed as follows:
\begin{eqnarray}
  \label{eq:amp-3}
  A^{b\to s}\!\!\!\!&=& e^{i\Theta}\big[ a+b\,e^{i\delta}
  e^{i\gamma}\big],\\
  \label{eq:a_t} a&=&\left|v_c\right|\hat{a}=\left|v_c\right|\left|{\cal A}_c
    e^{i\delta_c}-{\cal A}_t 
    e^{i\delta_t}\right|,\\
  \label{eq:b_t}
  b&=&\left|v_u\right|\hat{b}=\left|v_u\right|\,\left|{\cal A}_u    
    e^{i\delta_u}-{\cal A}_t e^{i\delta_t}\right|,
\end{eqnarray}
$\delta$ is the strong phase difference between $a$ and $b$; $\Theta$
is the overall strong phase which is set to zero, as no observables
depend on it.  The same amplitude could have been written in terms of
$\beta_s$ by eliminating $v_u$. However,
it is not a priori necessary to choose the parameterization to start
with.  

The amplitude $A_i$ for $\bd$ decay to any mode $f_i$ can be
expressed in a parameterization independent way, in terms for two
contributing amplitudes as
\begin{equation}
  \label{eq:amps-3}
  {A}_i=a_i +b_i e^{i\delta_i} e^{i\phi}.
\end{equation}
$a_i$ and $b_i$ are the two amplitudes contributing to the
process and $\delta_i$ is the corresponding strong phases difference.
$\phi$ is the weak phase that could either be $\gamma$ or $\beta_s$ to
distinguish between the two parameterizations. By definition we choose
$a_i$ and $b_i$ to be the magnitudes with the relative phase
$\delta_i$ taking care of all relative signs. Note that $v_t$ has a
negative sign and $\delta_i$ is defined so as to include this sign
when the amplitude is expressed in terms of $\beta_s$.  Using CPT
invariance, the amplitude for the conjugate mode is given by
$\bar{A}_i=a_i+b_i e^{i\delta_i} e^{-i\phi}$.  To simplify notation
and without loss of generality, we assume that the amplitude ${A}_i$,
$a_i$ and $b_i$ are normalized by the total $B^0$ decay width. The
time dependent decay rate of $\bd$ to a mode $f_i$ using these
amplitudes may then be written as
\[ \Gamma(B^0(t)\!\to f_i) \propto B_i (1 + C_i\cos(\Delta\! M t) -
S_i\sin(\Delta\! M t)), \]
where $B_i$ is the branching ratio, $C_i$ is the direct $CP$ asymmetry
and $S_i$ is the time dependent asymmetry, given by
\begin{eqnarray}
  \label{eq:Bi}
  B_i\!&=&\!\frac{|{A}_{i}|^2+|\bar{
      A}_{i}|^2}{2}\!=\!a_i^2+b_i^2+2\,a_i\,b_i\, 
  \cos\phi\cos\delta_i, \nn \\
  \label{eq:Ci}
  C_i\!&=&\!\frac{|A_{i}|^2-|\bar{A}_{i}|^2}
  {|A_{i}|^2+|\bar{A}_{i}|^2}\!=\!  
  \frac{-2\,a_i\,b_i\,\sin\phi\sin\delta_i}{B_i}, \\
  \label{eq:Si}
  S_i\!&=&\!\sqrt{1-C_i^2}\sin 2\beta_i^{\meas}\!=\!-\sqrt{1-C_i^2}\,\frac{{\rm
      Im}(e^{-2i\beta} A_{i}^*{\bar
      {A}_i})}{|A_i||\bar{A}_i|}.\nn
\end{eqnarray}

The time dependent asymmetry provides a measurement of
$\sin(2\beta_i^{\meas})$, hence, $2\beta_i^{\meas}$ is obtained with a
two fold ambiguity $(2\beta_i^{\rm meas},\pi-2\beta_i^{\meas})$. This
in turn leads to a four fold ambiguity in the difference between the
values of $2\beta^{\meas}$, obtained using the two different modes
$f_1$ and $f_2$. However, {\em we will only be interested in the
  principal value} $2\beta_i^{\meas}$, obtained from
$\sin(2\beta_i^{\meas})$, so as to have a well defined value of the
difference.  We denote this value of the difference by $2 \omega$ and
define it as $2\omega=2\beta_1^{\meas}-2\beta_2^{\meas}$.  We assume
that the two modes $f_1$ and $f_2$ are chosen such that
$\beta_1^{\meas}>\beta_2^{\meas}$. This choice results in $2 \omega$
always being positive.  We further, define the phase difference
between ${A}_{i}$ and $\bar{A}_{i}$ as $\eta_i$, i.e.,
$\eta_i=\arg{{A}_{i}}-\arg{\bar{A}_{i}}$. Hence,
${A}_{i}^*{\bar{A}}_{i}=|{A}_{i}||\bar{A}_{i}| e^{-i\eta_i}$ and the
expression for $S_i$ from Eq.~(\ref{eq:Ci}) implies that
$\eta_i=2\beta_i^{\meas}-2\beta$. $\eta_i$ is thus the deviation of
$2\beta_i^{\meas}$ from $2\beta$. $\omega$ can now be expressed in
terms of $\eta_{1,2}$ as $2\omega=\eta_1-\eta_2$.

Only three independent observables $B_i$, $C_i$ and $S_i$ can ever be
obtained using the decay modes under consideration, but these
observables can be described in terms of five variables $a_i$, $b_i$,
$\delta_i$, $\beta$ and $\phi$. We can hence express three of these
variables in terms of observables and the other two variables.  We
will choose to express $a_i$, $b_i$ and $\delta_i$ in terms of $\beta$
and $\phi$.  It is easy to derive expressions for $a_i$, $b_i$ and
$\delta_i$ using Eqns.~(\ref{eq:amps-3}-\ref{eq:Ci})
\begin{eqnarray}
  \label{eq:a}
  a_i^2&=&\frac{B_i}{2\sin^2\phi}\Big(1-\sqrt{1-C_i^2}
  \cos(\eta_i-2\phi)\Big)  \\  \label{eq:b}   
  b_i^2&=&\frac{B_i}{2\sin^2\phi}\Big(1-\sqrt{1-C_i^2}
  \cos(\eta_i)\Big) \\ \label{eq:tan_delta}
  \tan\delta_i&=&\frac{C_i \sin\phi}{\cos\phi-\sqrt{1-C_i^2}\cos(\eta_i-\phi)}
\end{eqnarray}

Our interest is in finding a relation between $\omega$ and $\phi$.
Hence, we first find a relation between $\eta_i$ and $\phi$. Indeed,
Eq.~(\ref{eq:tan_delta}) expresses $\eta_i$ in terms of $\delta_i$ and
$\phi$. However, in this paper we present a geometric approach which
is much more intuitive though we have verified these results
numerically.
\begin{figure}[b]
\centering
\includegraphics*[scale=0.56]{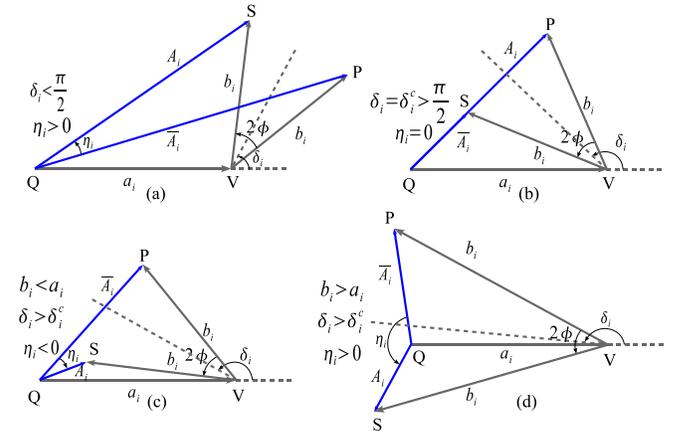}
\caption{\label{fig:sign_of_eta}The amplitudes $A_i$ and ${\bar A}_i$
  in terms of $a_i$ and $b_i$ for the case $\phi>0$ and $\delta_i>0$.} 
\end{figure}
In Fig.~\ref{fig:sign_of_eta} we have shown the variation of $\eta_i$
with respect to different values of $\delta_i$.  We begin by
first representing $A_i$ and $\bar{A}_i$ geometrically.  Given values
of $a_i$, $b_i$, $\delta_i$ and $\phi$, $|A_i|$ and $|\bar{A_i}|$ are
as shown in the Fig.~\ref{fig:sign_of_eta}. For the purpose of
illustration we have chosen $\delta_i>0$ and $\phi>0$. $\vec{a}$ is
represented by $QV$, and $\vec{b}$ is represented by $SV$ or $PV$
depending on the phase $\delta+\phi$ or $\delta-\phi$, resulting in
the amplitude $A$ and $\bar{A}$ respectively.  It may be noted that
the same values of $|A_i|$, $|\bar{A_i}|$ and $\eta_i$ can be
obtained using different values of $a_i$, $b_i$, $\delta_i$ and
$\phi$. The set of points for which this is possible is obtained by
moving the point $V$ along the bisector to $SP$, since $SV$ and $PV$
are both $b$, they must always be equal.  It is hence
essential to express all quantities in terms of irreducible variables.

In Fig.~\ref{fig:sign_of_eta}(a) we have chosen $\delta_i$ to lie in
the range between 0 and $\pi/2$. Clearly $\eta_i$ is always positive
(if $\phi>0$) irrespective of the value of amplitudes $a_i$ and $b_i$.
If $\delta_i$ is increased beyond $\pi/2$, at some critical value of
$\delta_i=\delta_i^c$, $\eta_i$ becomes 0 (see
Fig.~\ref{fig:sign_of_eta}(b)). If $\delta_i$ is increased further
beyond $\delta_i^c$ the sign of $\eta_i$ depends on the relative
magnitudes of $a_i$ and $b_i$; $\eta_i <0$ if $b_i<a_i$
(Fig.~\ref{fig:sign_of_eta}(c)) and $\eta_i >0$ if $b_i>a_i$
(Fig.~\ref{fig:sign_of_eta}(d)). It is easy to generalize to the cases
where both $\delta_i$ and $\phi$ can be positive or negative.  Note
that if $\delta_i$ and $\phi$ have the same sign then $C_i<0$, else
$C_i>0$. The value of the critical angle $\tan\delta_i^{c}$ is easily
obtained using Eq.~(\ref{eq:tan_delta}) by setting $\eta_i=0$.
$\delta_i^{c}$ lies in the range $\pi/2$ to $\pi$ for $C_i<0$
and $-\pi$ to $-\pi/2$ for $C_i>0$
\footnote{Fig.~\ref{fig:sign_of_eta}(b)
  and \ref{fig:sign_of_eta}(c) do not apply to the case where $C_i=0$,
  as only two values of $\delta_i$ are allowed, it can either be $0$
  (limiting case of Fig.~\ref{fig:sign_of_eta}(a)) with $\eta_i>0$, or
  $\pi$ (limiting case of Fig.~\ref{fig:sign_of_eta}(d) and
  \ref{fig:sign_of_eta}(e)) with $\eta_i >0$ if $b_i>a_i$ and
  $\eta_i<0$ if $b_i<a_i$.  $\delta_i^c$ is strictly not defined for
  $C_i=0$, but may be taken to have any value between $\pi/2$ and
  $\pi$.}.  Hence, {\em we conclude that $\eta_i$ always has the same
  sign as $\phi$ if $\left|\delta_i\right|\leq
  \left|\delta_i^c\right|$}.  The weak phase $\phi$ is fixed by the
parameterization chosen within SM and is the same for all modes.
Hence as long as $\left|\delta_i\right|\leq \left|\delta_i^c\right|$
for each mode $f_i$, the sign of $\phi$ and $\eta_i$ must be the same
for all modes.

\begin{figure}[t]
\centering
\includegraphics[scale=0.275]{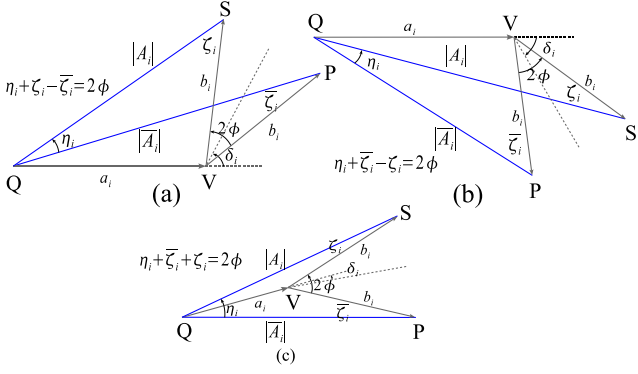}
\caption{\label{fig:2}Case $\phi>0$ and
  $-\delta_i^c<\delta_i<\delta_i^c$. The amplitudes $A_i$ and ${\bar
    A}_i$ in terms of $a_i$ and $b_i$.}
\end{figure}

Our next aim is to find a bound on the size of $\eta_i$ as a function
of $\phi$. Using Fig.~\ref{fig:2} we try to establish a bound on
$\eta_i$. We begin by choosing $\phi >0$.  However, the derivation of
the bound depends on the sign of $\delta_i$, and the two cases
$\delta_i<0$ and $\delta_i>0$ need to be considered separately. In
Fig.~\ref{fig:2}(a) we have chosen $0\leq\delta_i\leq \delta_i^c$. It
is easy to see that $2\phi=\eta_i+\zeta_i-\bar{\zeta}_i$.  Since both
$\phi>0$ and $\delta_i>0$, Eq.~(\ref{eq:Ci}) implies that $C_i <0$ or
$|\bar{A}_i|>|A_i|$ resulting in $\bar{\zeta}_i <\zeta_i$. We hence
conclude that $2\phi\geq \eta_i$. Fig.~\ref{fig:2}(b) depicts the case
for $\delta_i^c\leq\delta_i\leq 0$. In this case one has
$2\phi=\eta_i+\bar{\zeta}_i-\zeta_i$.  Since $\phi$ and $\delta_i$
have opposite signs, $C_i >0$. This implies that $|\bar{A}_i|<|A_i|$,
leading to the conclusion that $\bar{\zeta}_i > \zeta_i$ in this case.
The conclusion $2\phi\geq \eta_i$ obtained for Fig.~\ref{fig:2}(a)
still holds for Fig.~\ref{fig:2}(b). If the point $V$ lies inside
$\triangle SQP$ as shown in Fig.~\ref{fig:2}(c), one would obtain the
relation $ 2\phi=\eta_i + \bar{\zeta}_i + \zeta_i$, once again
implying that $2\phi\geq \eta_i$. Hence, in all the three cases
discussed in Fig.~\ref{fig:2}, one finds that $0\le\eta_i\le 2\phi$.
Within the SM, the weak phase $\phi$ is always positive
($\gamma\approx 60^o$ and $\beta_s\approx 1^o$). Hence we do not
consider the case for $\phi<0$ in detail but simply state that one can
straightforwardly obtain $\left |\eta_i\right| \le
2\left|\phi\right|$.

\begin{figure}[t]
\centering
\includegraphics[scale=0.35]{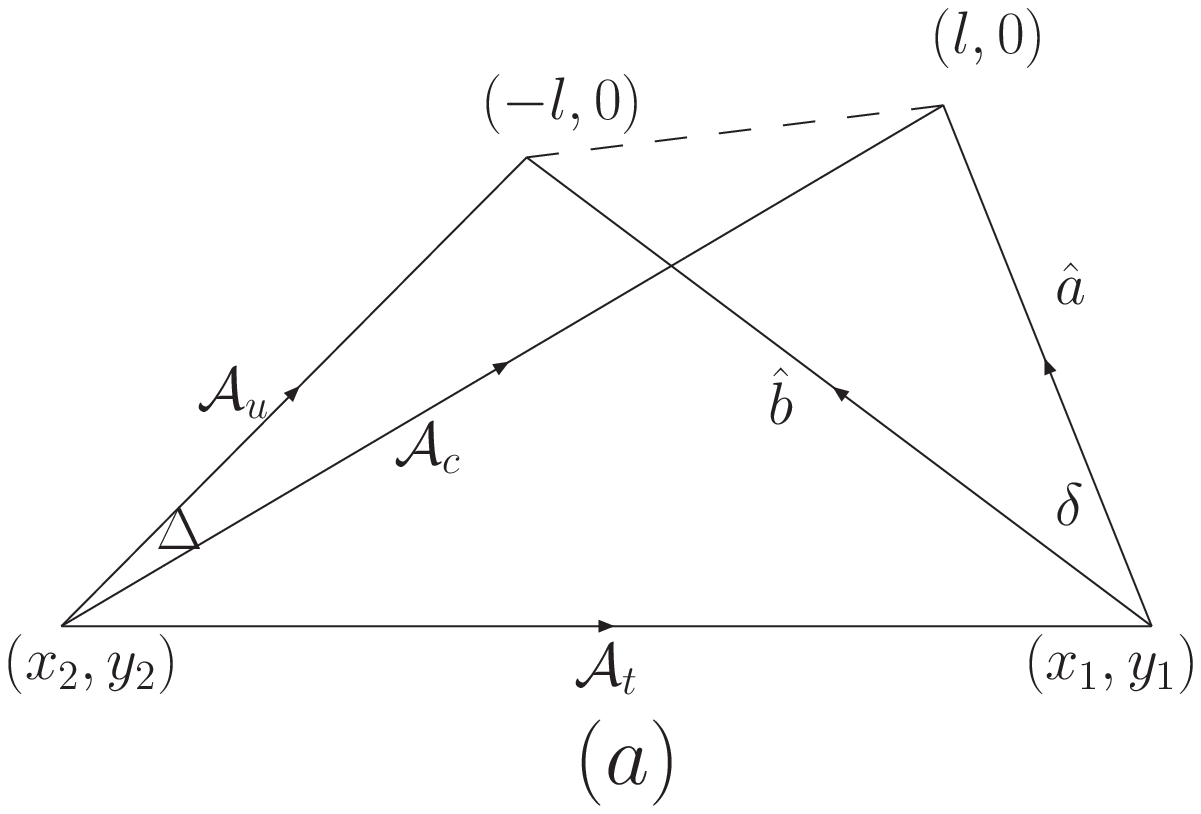}
\includegraphics[scale=0.35]{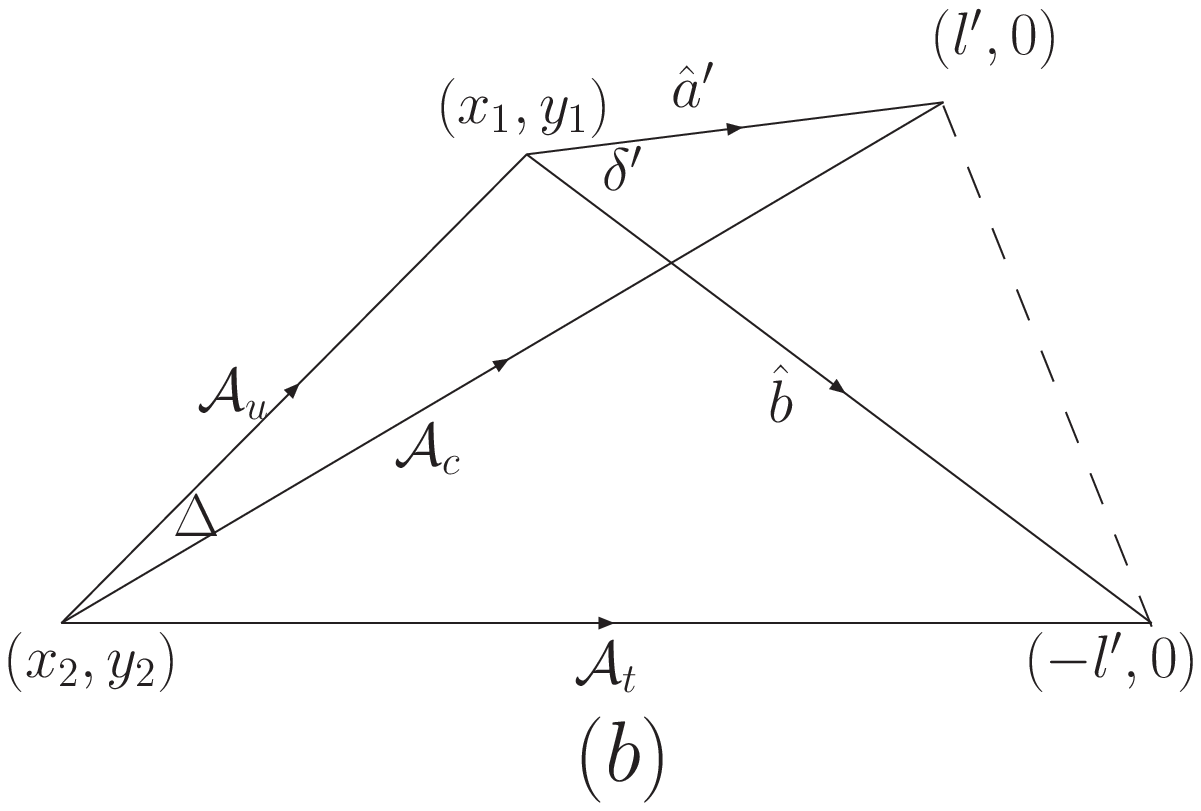}
\caption{\label{fig:3}Geometric representation of
  Eqns.~(\ref{eq:amp-3})-(\ref{eq:b_t}). In $(a)$ we choose the CKM
  parameterization with $\phi=\gamma$ and in $(b)$ with
  $\phi=\beta_s$. In $(b)$ a prime has been introduced only to
  distinguish variables that differ between the two
  parameterizations.}
\end{figure}

Note that while $\omega$ is an observable, $\eta_i$ itself is not an
observable, unless, $\beta$ is measured in some mode without
contribution from a polluting amplitude -- a theoretical
impossibility, even within the SM.  Nevertheless, it may be reasonable
to assume (as is customarily done) that $\mathcal{A}_c$ dominates in
$b\to c\bar{c}s$ indicating $\eta_1\approx 0$.  A contradiction is
already evident, since $\phi$ and hence $\eta_i$ must be positive in
the SM, implying that $2\omega\approx-\eta_2<0$.  The only way to
resolve the contradiction is to assume that 
$\left|\delta_i\right|\nleq\left|\delta_i^c\right|$.  Hence, before
drawing any conclusions we still need to determine when
$\left|\delta_i\right|>\left|\delta_i^c\right|$ or
$\left|\delta_i\right|<\left|\delta_i^c\right|$.  To this end, we note
that Eqns.~(\ref{eq:a_t}) and (\ref{eq:b_t}) can be graphically
represented as shown in Fig.~\ref{fig:3}. In Fig.~\ref{fig:3}(a) we
have chosen the CKM parameterization with $\phi=\gamma$ and in
Fig.~\ref{fig:3}(b) with $\phi=\beta_s$. It is easy to solve for the
coordinates $(x_1,y_1)$, $(x_2,y_2)$ and $l$ in terms of
$\mathcal{A}_{u,c}$ and $a_i$, $b_i$ and $\delta_i$. Since $a_i^2$,
$b_i^2$ and $\tan{\delta_i}$ can be evaluated as a function of
$\eta_i$ using Eqns.~(\ref{eq:a})-(\ref{eq:tan_delta}) purely in terms
of observables and $\phi$, one can express $\mathcal{A}_t$ and
$\Delta\equiv\left|\delta_c-\delta_u\right|$ in terms of
$\mathcal{A}_c$, $\mathcal{A}_u$ and $\eta_i$. Note that
$\mathcal{A}_j$ are CKM parameterization independent.

\begingroup
\squeezetable
\begin{table*}[hbt]
  \centering
    \begin{tabular}{|c|l|c|c|c|c|c|}
      \hline \hline \hline
      &&$0<2\phi$&$\eta_2$ bound&$\eta_1$ bound&$\beta$ bound&
      Constraints\\ [2ex]
      \hline
      $\beta_2^\meas\leq\beta_1^\meas\leq\beta$&I&$\eta_2\leq\eta_1\leq
      0\leq 2\phi$& $\eta_2\leq -2 \omega$ & $\eta_1\leq 0$&
      $2\beta_1^\meas\leq 2\beta$&$\eta_2\leq
      -13.63^o$\\[2ex]
      \hline
      $\beta_2^\meas\leq\beta\leq\beta_1^\meas$&II(a)&$\eta_2\leq
      0\leq\eta_1\leq 2\phi$& $\eta_2\leq 2\phi-2\omega$ &
      $0\leq \eta_1\leq 2\phi$&
      $2\beta_1^\meas-2\phi\leq 2\beta$&$\eta_2\leq
      -11.54^o$\\[2ex]
      \cline{2-7}&II(b)&
      $\eta_2\leq
      0\leq 2\phi\leq\eta_1$& $2\phi-2\omega\leq\eta_2$ & $2\phi\leq\eta_1\leq 2\omega$&
      $2\beta_2^\meas\leq 2\beta \leq 2\beta_1^\meas-2\phi$&$
      -11.54^o\leq\eta_2; 2.09^o \leq \eta_1 \leq 13.63^o$\\
      &&&&&&  $ 30^o \leq 2\beta \leq 41.54^o$\\[1.5ex]
      \hline
      $\beta\leq\beta_2^\meas\leq\beta_1^\meas$&III(a)&
      $0\leq\eta_2 \leq
      2\phi \leq\eta_1$& $0 \leq \eta_2\leq 2\phi$ & $2\omega\leq\eta_1$&
      $2\beta_2^\meas- 2\phi\leq 2\beta \leq
      2\beta_2^\meas$&$13.63^o\leq\eta_1$\\ 
      &&&&&&  $27.91^o \leq 2\beta \leq 30^o$\\[1.5ex]
      \cline{2-7}&III(b)&
      $0\leq 2\phi \leq \eta_2 \leq \eta_1
      $& $2\phi\leq\eta_2$ & $ 2\omega +2\phi\leq\eta_1$&
      $2\beta \leq 2\beta_2^\meas-2\phi$&$15.72^o\leq\eta_1
      ;2\beta \leq 27.91^o$\\[2ex]
      \hline
      \hline
      \hline
    \end{tabular}
    \caption{Constraints on $\eta_i$ and $2\beta$ for the
      parameterization $\phi=\beta_s$. Note that when $\eta_i<0$ or
      $\eta_i>2\phi$ one must have $|\delta_i|>|\delta_i^c|$. Note
      that in none of the cases it is possible to have
      $|\delta_i|<|\delta_i^c|$ for both modes. Unless
      $0<2\omega<2\phi$ one cannot have $0\leq\eta_2\leq\eta_1\leq
      2\phi$, hence we do not consider this specific case.}
  \label{table1}
\end{table*}
\endgroup
\begin{figure}[b]
\centering
\includegraphics[scale=0.45]{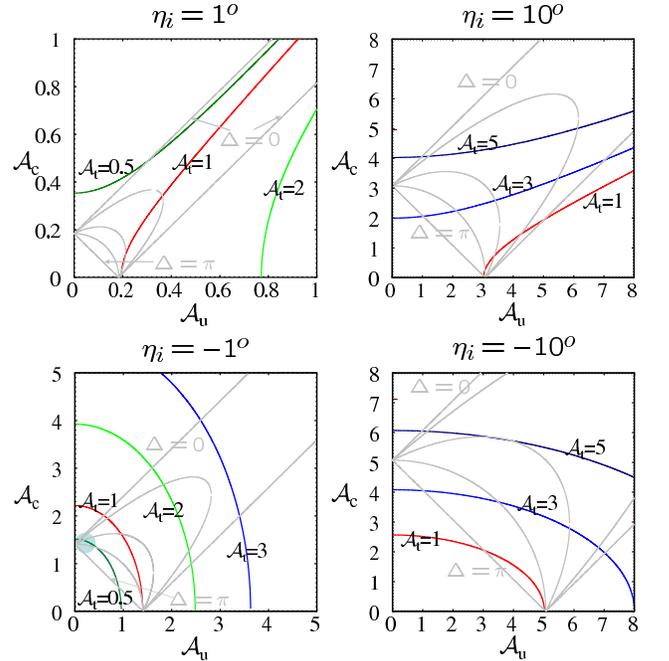}
\caption{\label{fig:4}Values of ${\cal A}_t$ and
  $\Delta\equiv\left|\delta_u-\delta_c\right|$ as a function of ${\cal
    A}_u$ and ${\cal A}_c$. ${\cal A}_j$ are normalized such that, if
  ${\cal A}_u=0$ and ${\cal A}_t=0$, ${\cal A}_c$ would be unity.  The
  allowed values are bounded by the curves for $\Delta=0,\pi$. The
  unlabeled parabolic curves represent $\Delta=\frac{\pi}{2}$,
  $\frac{\pi}{3}$ and $\frac{\pi}{6}$. }
\end{figure}

In Fig.~\ref{fig:4} we have plotted the values of $\mathcal{A}_t$ and
$\Delta$ in terms of $\mathcal{A}_u$ and $\mathcal{A}_c$. The observed
$C_i$'s are consistent with zero, hence, we have assumed $\delta_i=0$
for $0\leq\eta_i\leq 2\phi$ and $\delta_i=\pi$ for $\eta_i<0$ or
$\eta_i>2\phi$.  Non-zero values of $C_i$'s result in more stringent
constraints. This is easy to see from
Eqns.~(\ref{eq:a})-(\ref{eq:tan_delta}) and Fig.~\ref{fig:3}.  Finite
$C_i$ implies larger $a_i$, $b_i$ and $|\tan\delta_i|$ compared with
the $C_i=0$ values, this in turn results in larger $\mathcal{A}_j$.
$\mathcal{A}_j$ are normalized such that, if ${\cal A}_u=0$ and ${\cal
  A}_t=0$, ${\cal A}_c$ would be unity.  A clear pattern of hierarchy
of the various values of $\mathcal{A}_j$ becomes apparent which
depends only on $\eta_i$. Assuming it is reasonable for
$\mathcal{A}_c$ to dominate in $b\to c\bar{c}s$ decays, it is clear
from Fig.~\ref{fig:4} that only small negative values of $\eta_i$ are
allowed. This region is shown in the plot for $\eta_i=-1^o$ with a
small (blue) circle. 

The value of $\omega$ can be estimated from the available world
average values of two modes, however, as is customary, to improve
statistics we use the average values \cite{HFAG} of
$\sin2\beta^{\meas}$ measured using several $b \rightarrow c\bar{c}s$
and $b \rightarrow s\bar{q}q$ modes which are as follows: $
\sin2\beta_{1}^{\meas}(b \rightarrow c\bar{c}s)= 0.69 \pm 0.03$ and
$\sin2\beta_{2}^{\meas}(b \rightarrow s\bar{q}q) = 0.50\pm 0.06$,
implying $2\omega = (13.63\pm 5.41)^o$.  CKM fits \cite{CKMfit} on the
other hand give $\sin2\beta\approx 0.75$.

We now examine the possible constraints on the parameters
$\mathcal{A}_u$, $\mathcal{A}_c$, $\mathcal{A}_t$ and $\beta$ from
data without making any hadronic model assumptions.  There are three
possible values of $\beta$: it can either be greater than both
$\beta_1^\meas$ and $\beta_2^\meas$ or less than both of them or in
between both of them.  In Table~\ref{table1} we consider the cases
with the possible sub cases depending on the value of $\phi$ to obtain
bounds on $\eta_1$, $\eta_2$ and $2\beta$.  We could have chosen any
of the parameterization with $\phi$ being either $\gamma$ or
$\beta_s$.  However, {\em for the choice $\phi=\beta_s$, the bounds
  are more stringent}.  Hence, we have set $2\phi\approx 2^o$ in
Table~\ref{table1}. It is interesting that SM allows for such a small
value of $\beta_s$. {\em It is the smallness of this value that works in
our favor}.

The bounds on $\eta_i$ obtained in Table~\ref{table1} have
corresponding constraints on quark level amplitudes as shown in
Fig.~\ref{fig:4}. It may be noted that only for Case I of
Table~\ref{table1}, it is possible to have $\mathcal{A}_{t,u} <
\mathcal{A}_c$, which is expected in $b\to c\bar{c}s$ modes. Hence all
the cases except Case I are unrealistic. Further, in all the cases one
finds that the bounds on $\eta_2$ and (or) $\eta_1$ are such that,
they require the square of the quark level amplitudes
$|\mathcal{A}_c|^2$ ($|\mathcal{A}_u|^2$) and $|\mathcal{A}_t|^2$ that
are at least 10 to 25 times larger than the observed branching ratio.
The values close to 10 are possible only by lowering $2\beta$ away
from $2\beta_1^\meas$, beyond acceptable values, close to
$2\beta_2^\meas$.  Hence, none of the cases can be accommodated within
the SM, unless one requires that the observed branching ratios result
from considerable fine tuned cancellations of quark level amplitudes.
We wish to emphasize that even if $2 \omega$ reduces to $5^o$ in
future, it would still be difficult to accommodate within the SM.
Since one would still requires squares of the amplitudes
$|\mathcal{A}_j|^2$ that are 10 times larger than the observed
branching ratios.

To conclude, without making any hadronic model based assumptions we
have shown that within the SM, it is impossible to explain the
observed discrepancy in $\bd-\bdbar$ mixing phase measured using the
$b\to c\bar{c}s$ and $b\to s\bar{q}q$ modes.  The only possibility to
forgo this conclusion is to accept that the observed branching ratios
result from considerable fine tuned cancellations of significantly
larger quark level amplitudes.  In addition it may also be necessary
to have $2\beta$ substantially lower than $2\beta_1^\meas$. This
scenario of ``observed decay rates resulting from fine-tuned
cancellations of large quark level amplitudes'' would be very
difficult to accommodate, given the successful understanding of $B_d$
decay rates.

RS thanks the National Center for Theoretical Sciences at Taipei for
hospitality where part of this work was done. WSH thanks the Institute
of Mathematical Sciences for hospitality. The work of WSH was
supported in part by  NSC-93-2112-M-002-020 of Taiwan.

\end{document}